\documentclass{article}
\usepackage{spconf,amsmath,graphicx,hyperref}

\usepackage{float}
\usepackage{setspace}
\usepackage{tikz}
\usepackage{makecell}
\usepackage[table]{xcolor}
\usetikzlibrary{arrows.meta}
\usepackage{mathrsfs}
\usepackage{amsmath}
\usepackage{pgfplots}
\usepackage{enumitem}
\pgfplotsset{compat=1.18}
\usepackage{booktabs}
\usepackage{subcaption}
\usepackage{threeparttable}
\usepackage{placeins}

\usepackage{pgf}  
\usepackage{cite}
\usetikzlibrary{calc} 
\usetikzlibrary{positioning}
\newcommand{\tagabove}[2]{\begin{tabular}[c]{@{}c@{}}\bfseries(#1)\\ #2\end{tabular}}

\usepackage{fancyhdr}

\fancypagestyle{firstpage}{
    \fancyhf{}
    \fancyhead[L]{Accepted to 2026 IEEE International Conference on Acoustics, Speech, and Signal Processing (ICASSP)} 
}

\usepackage{balance}

\definecolor{myCream}{RGB}{255,253,208} 
\definecolor{myPurple}{RGB}{250,240,250}   
\definecolor{myGreen}{RGB}{230,250,230}    
\colorlet{myPurpleFill}{pink!20!white}
\colorlet{myLighterGreen}{myGreen!60!white} 
\definecolor{myLightBlue}{RGB}{179,229,252}
\colorlet{myBlue}{myLightBlue!60!white}
\definecolor{myPeach}{RGB}{255,229,204} 
\definecolor{OIskyblue}  {HTML}{56B4E9}

\makeatletter
\let\oldthebibliography\thebibliography
\let\endoldthebibliography\endthebibliography
\renewenvironment{thebibliography}[1]{%
  \oldthebibliography{#1}%
  \setlength{\itemsep}{0pt}%
  \setlength{\parskip}{0pt}%
  \setlength{\parsep}{0pt}%
}{\endoldthebibliography}
\makeatother

\title{Enhancing Value Alignment of LLMs with Multi-agent system \\ and Combinatorial Fusion}
\name{Yuanhong Wu$^{\star}$, Djallel Bouneffouf$^{ \dagger}$, and D. Frank Hsu$^{\star}$}
  
  \address{$^{\star}$  Dept. of Computer and Information Science, Fordham
        University, New York, NY, USA \\
      $^{\dagger}$ T. J.  Watson Research Center, IBM Research, Yorktown  
        Heights, NY, USA}

\begin{document}
\thispagestyle{firstpage}
%
\maketitle
\begin{abstract}
Aligning large language models (LLMs) with human values is a central challenge for ensuring trustworthy and safe deployment. While existing methods such as Reinforcement Learning from Human Feedback (RLHF) and its variants have improved alignment, they often rely on a single evaluator or narrowly defined reward signals, limiting their ability to capture ethical pluralism. In this work, we propose the Value Alignment System using Combinatorial Fusion Analysis (VAS-CFA), a framework that operationalizes multi-agent fusion alignment. It instantiates multiple moral agents, each fine-tuned to represent a distinct normative perspective, and fuses their outputs using CFA with both rank- and score-based aggregation. This design leverages cognitive diversity, between agents, to mitigate conflicts and redundancies across multiple agents, producing responses that better reflect human values. Empirical evaluation demonstrates that VAS-CFA outperforms both single agent baselines and prior aggregation approaches on standard metrics, showing that multi-agent fusion provides a robust and effective mechanism for advancing value alignment in LLMs.
\end{abstract}
\section{Introduction}
\label{sec:intro}

Aligning large language models (LLMs) with human values is critical because models pretrained on broad web corpora can produce outputs that are untruthful, unsafe, or misaligned with user intentions \cite{amodei2016concrete,wang2023aligning}. Alignment methods were developed to close these gaps.  In recent years, numerous techniques have been developed to better align LLMs with human values, several of which we summarize below.

The canonical alignment approach is Reinforcement Learning from Human Feedback (RLHF). This process involves supervised fine-tuning on demonstrations, followed by policy optimization using a reward model trained from human pairwise preferences, yielding strong improvements in toxic or unsafe outputs \cite{ouyang2022training}. To reduce the high cost of human labeling, RLAIF \cite{bai2022constitutional} largely replaces human raters with AI judges guided by a written constitution that supplies critiques and preference labels, producing a harmless but not evasive assistant with far fewer human ratings. Other methods aim to simplify the complex optimization process. For instance, Direct Preference Optimization (DPO) \cite{rafailov2023direct} replaces online RL with a simple supervised objective over preference pairs, yet still matches or exceeds the performance of RLHF with proximal policy optimization (PPO). 


\begin{figure*}[h!]
    \centering
    \resizebox{0.97\linewidth}{!}{%
\begin{tikzpicture}[node distance=2.5cm, scale = 1]

\node[] at (-0.9, 0) {\tiny{Agent A}};
\node[] (a1) at (0, 0){\includegraphics[width=0.8cm]{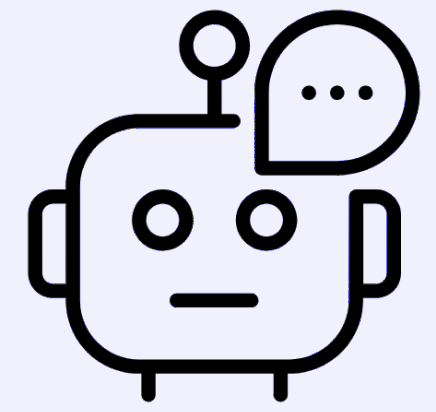}};
\node[draw, rectangle, right of=a1,xshift = -0.7cm, rounded corners = 5pt, fill = myCream] (b1) {a resp.};

\node[] at (-0.9, -0.8) {\tiny{Agent B }};
\node[] (a2) at (0, -0.8){\includegraphics[width=0.8cm]{agent.png}};
\node[draw, rectangle, right of=a2,xshift = -0.7cm,  rounded corners = 5pt , fill = myCream] (b2) {a resp.};

\node[] at (-0.9, -1.6) {\tiny{Agent C}};
\node[] (a3) at (0, -1.6){\includegraphics[width=0.8cm]{agent.png}};
\node[draw, rectangle, right of=a3,xshift = -0.7cm,  rounded corners = 5pt, fill = myCream] (b3) {a resp.};

\node[] at (-0.9, -2.4) {\tiny{Agent D}};
\node[] (a4) at (0, -2.4){\includegraphics[width=0.8cm]{agent.png}};
\node[draw, rectangle, right of=a4, xshift = -0.7cm, rounded corners = 5pt, fill = myCream] (b4) {a resp.};

\node[] at (-0.9, -3.2) {\tiny{Agent E}};
\node[] (a5) at (0, -3.2){\includegraphics[width=0.8cm]{agent.png}};
\node[draw, rectangle, right of=a5,xshift = -0.7cm, rounded corners = 5pt, fill = myCream] (b5) {a resp.};

\draw[->, -{Stealth}] (a1) -- (b1) node[midway, above] {\tiny{Inference}};
\draw[->, -{Stealth}] (a2) -- (b2) node[midway, above] {\tiny{Inference}};
\draw[->, -{Stealth}] (a3) -- (b3)  node[midway, above] {\tiny{Inference}};
\draw[->, -{Stealth}] (a4) -- (b4)  node[midway, above] {\tiny{Inference}};
\draw[->, -{Stealth}] (a5) -- (b5)  node[midway, above] {\tiny{Inference}};

\node[draw, rounded corners = 5pt, fill = myPurpleFill] (u1) at (5.5, 0) {unit 1};
\node[draw, rounded corners = 5pt, fill = myPurpleFill] (u2)at (5.5, -0.8) {unit 2};
\node[draw, rounded corners = 5pt, fill = myPurpleFill] (u3)at (5.5, -1.6) {unit 3};
\node[] at (5.5, -2.4) (u4) {$\vdots$};
\node[draw, rounded corners = 5pt, fill = myPurpleFill] (u5)at (5.5, -3.2) {unit $n$};

\node[draw, rounded corners = 5pt, fill=myPurpleFill] (pool) at (3.6, -1.5) {\makecell{moral \\ unit \\ pool}};
\draw[->, -{Stealth}] (b1.east) to[out=0, in=180, looseness=0.9] (pool.west);
\draw[->, -{Stealth}] (b2.east) to[out=-10, in = 180, looseness=0.9] (pool.west);
\draw [->, -{Stealth}] (b3.east) to[out=0, in = 180, looseness = 0.9] (pool.west);
\draw[->, -{Stealth}] (b4.east) to[out=0, in = 180, looseness=0.9] (pool.west);
\draw[->, -{Stealth}] (b5.east) to[out=0, in=180, looseness=0.9] (pool.west);

\draw[->, -{Stealth}] (pool.east) to[out=0, in=180, looseness=0.9] (u1.west);
\draw[->, -{Stealth}] (pool.east) to[out=-10, in = 180, looseness=0.9] (u2.west);
\draw [->, -{Stealth}] (pool.east) to[out=0, in = 180, looseness = 0.9] (u3.west);
\draw[->, -{Stealth}] (pool.east) to[out=0, in = 180, looseness=0.9] ($(u4.west)+(-0.3cm, 0)$);
\draw[->, -{Stealth}] (pool.east) to[out=0, in=180, looseness=0.9] (u5.west);

\node[] (c) at (9.6, -0.5){\includegraphics[width=5.5cm]{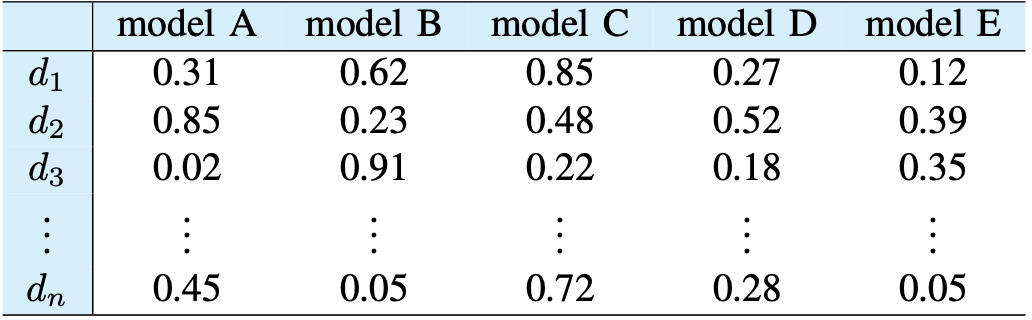}};

\node[] (temp1) at (7.1, -0.07) {};
\node[] (temp2) at (7.1, -0.28) {};
\node[] (temp3) at (7.1, -0.48) {};
\node[] (temp4) at (7.1, -0.8) {};
\node[] (temp5) at (7.1, -1.15) {};
\draw[->,  -{Stealth}] (u1.east) -- (temp1); 
\draw[->, -{Stealth}] (u2.east) -- (temp2);
\draw[->, -{Stealth}] (u3.east) -- (temp3);
\draw[->, -{Stealth}] ($(u4.east)+(0.35cm, 0)$) -- (temp4);
\draw[->, -{Stealth}] (u5.east) -- (temp5);

\node[draw, rounded corners = 5pt, fill = myPeach] (d1) at (8.5, -2) {score comb.};
\node[draw, rounded corners = 5pt, fill = myPeach] (d2) at (11, -2) {rank comb.};

\node[draw, rounded corners = 5pt, fill = myPeach] (e1) at (7.8, -3.2) {\makecell{average comb.\\ \scriptsize{(ASC/ARC)}}};
\node[draw, rounded corners = 5pt, fill = myPeach] (e2) at (11.4, -3.2) {\makecell{weighted comb.\\ \scriptsize{(WSCDS/WRCDS)}}};

\draw[->, -{Stealth}] (d1) -- (e1);
\draw[->, -{Stealth}] (d1) -- (e2);

\draw[->, -{Stealth}] (d2) -- (e1);
\draw[->, -{Stealth}] (d2) -- (e2);

\draw (9.8, -1.36) -- (9.8, -1.6)  node[midway, right] {}; 
\draw[->, -{Stealth}] (9.8, -1.6) -- (8.8, -1.6) -- (8.8, -1.8);
\draw[->, -{Stealth}] (9.8, -1.6) -- (10.8, -1.6) -- (10.8, -1.8);

\node[
  draw,
  rounded corners=5pt,
  fill=myLighterGreen,
  align=left
] at (14.5, -1.5) {%
Paraphraser\\
{\small $\bullet$ Select best model}\\
{\small $\bullet$ Paraphrase top unit}
};


\draw[->, -{Stealth}] (9.9, -3.9) -- (9.9, -4.1) -- (14.4, -4.1)  node[pos=0.83, above] {} -- (14.4, -2.5);

\draw[line width=0.5pt] (8, -3.71) -- (8, -3.9) -- (11.7, -3.9) -- (11.7, -3.71);

\draw[dotted, very thick, blue] (6.56,0.5) rectangle (12.8,-4.2);
\node[] at (7.6, 0.8) {(c) CFA step};
\node[] at (1.7, 0.8) {(a)};
\node[] at(5.4, 0.8) {(b)};
\node[] at(13.7, 0.8) {(d)};
\end{tikzpicture}
}
    \caption{The diagram for Value Alignment System using Combinatorial Fusion Analysis (VAS-CFA).}
    \label{fig:diagram}
\end{figure*}

Moving beyond pairwise preferences, researchers are now leveraging richer textual signals to directly steer model behavior. For instance, Self-Refine \cite{madaan2023self} instructs a model to revise its own output, improving both human preference and task performance scores without any additional weight updates. Similarly, Reflexion \cite{shinn2023reflexion} converts feedback into verbal self-supervision by having the model generate brief reflections that guide its subsequent attempts, yielding better reasoning  performance without additional training. Unifying these concepts, Diverse AI Feedback (DAIF) \cite{yu2025diverse} combines critique, refinement, and preference into a single pipeline. It routes tasks to the most informative feedback type, outperforming single-feedback baselines. This demonstrates that heterogeneous feedback is a powerful mechanism for improving alignment.

A growing literature warns that RLxF can result in narrow objectives and miss crucial ethical complexity \cite{dahlgren2025helpful}. In response, methods are emerging to address this gap. User-Driven Value Alignment \cite{fan2025user} documents users’ strategies to notice, contest, and correct biased outputs, motivating interfaces for bottom-up steering and collective auditing. At community scale, STELA \cite{bergman2024stela} conducts deliberative norm-elicitation with underrepresented groups and derives a community ruleset for alignment, positioning alignment criteria as iterative and revisable.





The system we proposed in this paper contrasts with prior alignment methods centered on a single agent. The Value Alignment System using CFA (VAS-CFA) assembles multiple, diverse moral agents, each instantiated to capture a distinct normative perspective, and fuses their signals via a combinatorial fusion analysis framework with score-based and rank-based aggregation. By leveraging diversity across agents rather than relying on a single evaluator, and by explicitly integrating rankings into the fusion step, VAS-CFA offers a new mechanism for steering LLMs toward behavior that better reflects human values.

\section{Combinatorial fusion analysis (CFA) and Kemeny Rank Space ($K_n$)}
\label{sec: CFA}

Combinatorial Fusion Analysis (CFA) provides methods and workflows for combining multiple scoring systems (MSS) (e.g., multiple classifier systems, multiple expert systems, multiple language models, and multiple agent systems) in computational learning and modeling, informatics, and intelligent ML/AI systems \cite{hsu2006combinatorial, hsu2024combinatorial}. CFA characterizes a scoring system $A$ with a score function $s_A$, a derived rank function $r_A$, and a function that relates normalized score values to rank values \cite{hsu2006combinatorial,hsu2024combinatorial,hsu2010rank}.

\begin{figure}[h!]
    \centering
    \resizebox{\linewidth}{!}{\input{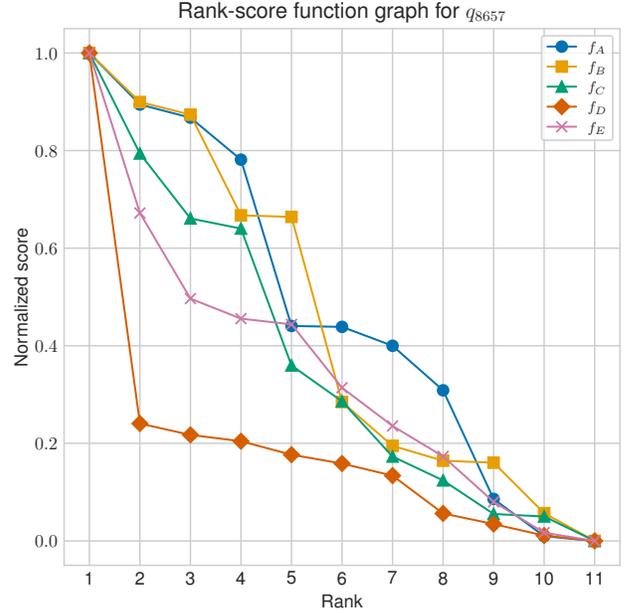}}
    \vspace{-0.5cm} 
    \caption{Rank-score function graph for the question $q_{8657}$: $f_A, f_B, f_C, f_D$ and $f_E$ refer to agent A, B, C, D and E w.r.t. Authority, Care, Fairness, Loyalty and Sanctity, respectively.}
    \label{fig:rsc}
\end{figure}

Let \(A\) be a scoring system on the dataset \(D=\{d_1,\ldots,d_n\}\). Let \(s_A:D\to R\) be a score function. Rank function \(r_A:D\to N\) is derived by sorting the score values and assigning an increasing rank value to the data item in \(D\) on the decreasing score values. Rank–score function (RSF) \(f_A:N\to R\) is defined as \(f_A(i)=s_A(r_A^{-1}(i))=(s_A \circ r_A^{-1} )(i)\). For scoring systems \(A\) and \(B\), cognitive diversity (CD) between $A$ and $B$, \(CD(A,B)\), is defined as the difference between $f_A$ and $f_B$ \cite{hsu2006combinatorial, hsu2010rank, hsu2019cognitive}: \(CD(A,B)=d(f_A,f_B)=\left(\sum_{i=1}^{n}(f_A(i)-f_B(i))^2/(n-1)\right)^{1/2}\). In addition, the diversity strength of the scoring system \(A\), \(DS(A)\), is the average of \(CD(A,A')\) between $A$ and all other scoring system \(A'\) under consideration. 

\begin{figure*}[t]
    \centering
    \resizebox{0.8\linewidth}{!}{\input{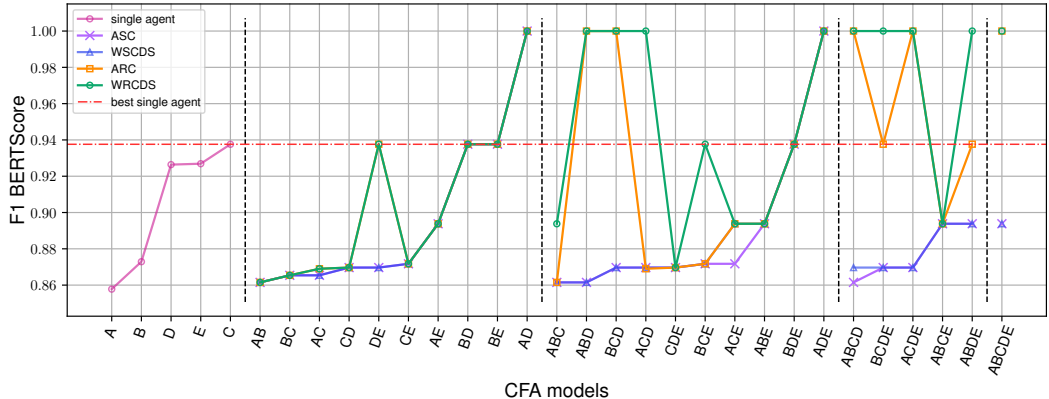}}
    \caption{F1 BERTScore across 26 combinations under four CFA combination types (ASC, WSCDS, ARC, WRCDS) for the question $q_{8657}$ (ASC sorted in non-decreasing order in each model group).}
    \label{fig:line_plot}
\end{figure*}

The weighted score combination by diversity strength (WSCDS) and the weighted rank combination by diversity strength (WRCDS) are defined for each $t$ in \{2, 3, 4, 5\} as below, respectively
{\small
$$s_{WSCDS}(d_i) = \frac{\sum_{A_j\in \mathcal{A}^\prime}s_{A_j}(d_i)\, \cdot\, DS(A_j)}{\sum_{A_j\in \mathcal{A}^\prime} DS(A_j)}, d_i \in D$$ 
$$s_{WRCDS}(d_i) = \frac{\sum_{A_j\in\mathcal{A}^\prime}r_{A_j}(d_i)\, \cdot\, \frac{1}{DS(A_j)}}{\sum_{A_j\in \mathcal{A}^\prime} \frac{1}{DS(A_j)}}, d_i \in D$$}where $\mathcal{A}^\prime$ is any subset of the full scoring system set $\mathcal{A} = \{A_1, \cdots, A_5\}$ containing at least 2 elements, i.e., $\mathcal{A}^\prime \subseteq \mathcal{A}$, and $\lVert \mathcal{A}^\prime \rVert = t$, with $t \in \{2,3, 4, 5\}$,  $s_{A_j}(d_i)$ and $r_{A_j}(d_i)$ are the score and rank assigned by scoring system $A_j$ to the data item $d_i$, respectively; $DS(A_j)$ represents the diversity strength of scoring system $A_j$.

 CFA has been applied to a wide variety of domain applications in scientific discovery and decision making, including more recent results in drug discovery and material science \cite{tang2021improving, yang2005consensus}. it was shown \cite{frank2005comparing} that rank combination tends to perform better w.r.t. larger cognitive diversity.

The CFA architecture entails both the continuous Euclidean space and a discrete rank space, corresponding to score function vectors and rank function vectors, respectively. When the score function is an 1-1 function, the rank function on the dataset \(D\) with \(n\) elements is a complete permutation of the \(n\) positive integers,
\([1,n]=(1,2, 3, \ldots,n)\). In this case, the set of complete permutations on the set \([1,n]\) is the symmetric group of order \(n\) under the composition operator (\cite{diaconis1988group} and \cite{jiang2023enhancing} suppl. inf.), denoted \(S_n\).
If we consider each permutation as a vertex in the graph of \(n!\) vertices and define adjacency between two vertices as a swap between an adjacent pair, the resulting graph is the bubble-sort Cayley graph \(B_n\).
(This Cayley graph is generated by a subset of vertices consisting of all transpositions between adjacent pairs of vertices). The graph \(B_n\) is \((n-1)\)-regular with connectivity \(n-1\), and \(B_n\) can be recursively constructed  from \(n\) copies of \(B_{n-1}\) \cite{jiang2023enhancing, zhong2019combining}.

When the score function is not 1-1, the derived ranking has ties. In this case, the ranking is not a complete permutation. Kemeny and Snell proposed a metric which includes tie ranking \cite{kemeny1962mathematical}. The Kemeny rank space \(K_n\) with the metric has number of vertices greater than \(n!\) (which is the number of vertices for \(B_n\)).
Works related to \(K_n\) include rank aggregation, multi-layer combinatorial fusion, and priority ranking \cite{akbari2023beyond, zhong2019combining, cook1978priority}.


\section{VAS-CFA workflow}
We propose a new framework that integrates Combinatorial Fusion Analysis into multiagent aggregation. We refer it to Value Alignment System using Combinatorial Fusion Analysis (VAS-CFA). Figure \ref{fig:diagram} shows the main steps for the proposed system.
In the first step of our VAS-CFA workflow, we fine-tuned five value-specific agents--Authority (A), Care (B), Fairness (C), Loyalty (D), and Sanctity (E)--starting from the OpenAssistant/oasst-sft-4-pythia-12b-epoch-3.5 SFT checkpoint, using Direct Preference Optimization (DPO) with QLoRA \cite{dettmers2023qlora} on a single NVIDIA A100-40GB. The Moral Integrity Corpus (MIC) \cite{ziems2022moral} is used for fine-tuning the individual agents. MIC dataset is publicly available and provides a large set of prompt-response pairs of 113.8K with human revised answers and rich ethical annotations. It uses fixed data splits of 91.0K/11.4K/11.4K samples for train/validation/test. Each moral agent was trained independently from the same base for 1 epoch with $\beta=0.1$, learning rate $1\times10^{-5}$, per-device batch size $2$ and $8$ gradient-accumulation steps. QLoRA loads the 12B base in 4-bit NF4 and trains only LoRA adapters, enabling practical single-GPU fine-tuning; We chose DPO because it optimizes human preferences without a reward model or online RL, which simplifies the pipeline and improves training stability and sample efficiency relative to PPO-based RLHF.


After fine-tuning five moral agents—each aligned with a distinct moral value—we obtain, for every test prompt, one generated response from each agent (Figure \ref{fig:diagram}(a)).  A naïve approach is to aggregate these five responses, following the value-fusion strategy of Dognin et al. \cite{dognin2025contextual}, which combines outputs using contextual information. However, direct aggregation risks semantic conflict: different agents may express incompatible moral commitments, yielding diluted or incoherent answers and weakening value alignment. Motivated by this, we decompose each agent’s output into \emph{moral units} using GPT-4.1 nano, where each unit conveys a single moral claim (Figure \ref{fig:diagram}(b)). This design is further supported by the observation that human-revised ground-truth answers are typically brief and contain only one or two moral ideas, whereas conversational model outputs are longer and multi-thematic. One example of breaking down a response to units is that Agent B (care) produces a response ``Promoting intelligence should be prioritized to ensure your child grows up healthy and prosperous", which is decomposed to three units "Promoting intelligence should be prioritized", ``Ensuring your child grows up healthy is important" and ``Ensuring your child grows up prosperous is important".



We then pool together all moral units extracted from the five agent responses. To score each unit for its alignment within each of the five moral values, we train a ``moral classifier.'' Concretely, we encode the human-revised answer using SentenceTransformer (all-MiniLM-L6-v2) and fit a logistic regression model for multi-label prediction. Given a moral unit, the classifier returns five scores, one per moral value, yielding the scoring table used in the CFA step (Figure \ref{fig:diagram}(c)). Interpreting each column as a value-specific scoring system, we obtain five scoring systems w.r.t. the five models A, B, C, D, and E for the five moral agents, respectively.


With the five scoring systems A, B, C, D, and E representing the five base moral agents, we use CFA described in Section \ref{sec: CFA} to produce $\sum_{i=1}^5 \binom{5}{i}=26$ combinations. In each of the 26 combinations, the CFA framework using diversity strength as weight, gives rise to four types of combinations: (1) average score combination (ASC), (2) weighted score combination by diversity strength (WSCDS), (3) average rank combination (ARC), and (4) weighted rank combination by diversity strength (WRCDS).

Next, we compare these 26 candidates, in one of the four types of CFA combinations (1)-(4), against the human-revised answer and retain the single best unit per configuration, yielding four units per prompt. In this study we proceed with the top unit only and pass it to a paraphraser that is prompted to answer the user's question while preserving the unit's moral content (Figure \ref{fig:diagram}(d)). This step is necessary because a moral unit encodes a moral claim, not necessarily a complete answer. 
If multiple units were selected, an aggregation module would instead fuse them into a coherent answer.
\looseness=-1

\section{Results}
\label{sec: results}

In this section, we report results for the VAS-CFA framework. 



Figure~\ref{fig:rsc} plots the rank-score function graphs for question $q_{8657}$ within test dataset of 113.8k prompt-response pairs. For each question, the five value-specific scoring systems yield five curves, which we overlay to visualize their mutual variation. Our results show that the five agents exhibit notable cognitive diversity across most of the questions in the test dataset.


\begin{table}[h]
\centering
\scriptsize
\begin{threeparttable}
\caption{Performance summary (ROUGE-L and $F_1$ BERTScore).}
\label{tab:agg_results}
\begin{tabular}{p{3cm}p{2cm}p{2cm}}
\toprule
\textbf{Model} &\tagabove{a}{\textbf{$F_1$ ROUGE-L}}&\tagabove{b}{\textbf{$F_1$ BERTScore}}  \\
\midrule
\multicolumn{3}{l}{(i). Individual moral agent}\\
 A & 0.0925 & 0.8569   \\ 
 B   & 0.0821  & 0.8533   \\ 
 C  &0.1249  & 0.8628  \\ 
 D  & 0.1376 & 0.8663   \\ 
 E  & 0.1343& 0.8653   \\ 

\hline
\addlinespace

\multicolumn{2}{l}{(ii). Fusion methods with CFA}\\
VAS-CFA: ASC & 0.1594 & 0.8831 \\ 
VAS-CFA: ARC & 0.1691 & 0.8847  \\ 
VAS-CFA: WSCDS  & 0.1598  & 0.8832  \\
VAS-CFA: WRCDS  & \textbf{0.1692}   & \textbf{0.8849} \\ 

\hline
\addlinespace

\multicolumn{2}{l}{(iii). Fusion methods without CFA}\\
Raw aggregation$^{\ast}$ & 0.1318  & 0.8654 \\
CVA-GS \cite{dognin2025contextual} & 0.1120 & 0.8728  \\
CVA-GS-DYN \cite{dognin2025contextual} & 0.1450 & 0.8754  \\ 
\bottomrule
\end{tabular}
\begin{tablenotes}[para,flushleft]
\footnotesize
\textit{Note:} Agents A, B, C, D and E refer to Authority, Care, Fairness, Loyalty and Sanctity, respectively. $^{\ast}$ Raw aggregation = aggregating the original five responses from the moral agents.
\end{tablenotes}
\end{threeparttable}
\end{table}

Results of the 26 combinations in Figure \ref{fig:line_plot} from question $q_{8657}$ show that rank combinations (ARC/WRCDS) consistently outperform score combinations (ASC/WSCDS)
due to cognitive diversity between moral agents as exhibited in Figure \ref{fig:rsc} \cite{frank2005comparing}.  Moreover, diversity strength (DS) as weight gives rise to desirable non-linear combination among individual agents.

Next, we evaluate VAS-CFA using the ROUGE\mbox{-}L metric. ROUGE\mbox{-}L measures the longest common subsequence between a system output and a reference, capturing sentence\mbox{-}level overlap while allowing for non\mbox{-}contiguous matches; we report the standard \(F_1\) score, which balances precision and recall against the human\mbox{-}revised answer. Table \ref{tab:agg_results}(a) summarizes the results across three groups of models. Group (i) consists of the five based models. Group (ii) lists results from each of four combinations (ASC, ARC, WSCDS, WRCDS). The group (iii) consists of raw aggregation, CVA-GS, and CVA-GS-DYN \cite{dognin2025contextual}.



Finally, Table \ref{tab:agg_results}(b) summarizes $F_1$ BERTScore results across three groups of models (i), (ii), and (iii) as in Table \ref{tab:agg_results}(a) for the F1 ROUGE-L score.

In both cases (a) $F_1$ ROUGE-L and (b) $F_1$ BERTScore in Table \ref{tab:agg_results}, VAS-CFA results exhibits improvement
 over results from single moral agents. It also outperforms 
previous multi-agent results by CVA-GS and CVA-GS-DYN. 
In addition, rank combinations (ARC/WRCDS) outperform 
score combinations (ASC/WSCDS) due to cognitive
 diversity between agents \cite{frank2005comparing}.

\section{Conclusion}
\label{sec: conclusion}
In this paper, we introduced Value Alignment System using Combinatorial Fusion Analysis (VAS-CFA), a framework that advances value alignment by operationalizing multiagent aggregation. Unlike prior approaches that rely on a single evaluator or narrowly defined reward signals, VAS-CFA instantiates multiple moral agents, each aligned with a distinct normative perspective, and integrates their outputs through Combinatorial Fusion Analysis (CFA). This design explicitly leverages cognitive diversity, using both rank- and score-based fusion to mitigate redundancy, resolve conflicts, and produce more coherent, value-sensitive responses.
Our experimental results demonstrate that the VAS-CFA is robust and consistently outperforms single-agent models and existing aggregation baselines across standard evaluation metrics. These findings highlight the potential of multiagent fusion as a powerful mechanism for capturing pluralistic values and improving alignment quality.

\small
\bibliographystyle{IEEEbib}
\bibliography{strings}



\end{document}